\immediate\write18{makeindex \jobname.nlo -s nomencl.ist -o \jobname.nls}

\documentclass[12pt]{article}

\usepackage[letterpaper, margin=1in]{geometry}

\usepackage{graphicx}
\usepackage[affil-it]{authblk}
\usepackage{float}
\usepackage{caption}
\usepackage{subcaption}
\usepackage{siunitx}

\usepackage{pgfplotstable}
\usepackage{tikz}
\usetikzlibrary{arrows,snakes,backgrounds,patterns,shapes.geometric,calc,shapes}

\usepackage{verbatim}
\usepackage{amsmath}
\usepackage{booktabs}
\usepackage{multicol}
\usepackage{nomencl}
\makenomenclature

\usepackage[colorlinks = true,
            linkcolor = blue,
            urlcolor  = blue,
            citecolor = blue,
            anchorcolor = blue]{hyperref}

\begin{document}

\title{Non-Linear Enthalpy Transformation for Transient Convective Phase Change in Smoothed Particle Hydrodynamics (SPH)}

\author{Amirsaman Farrokhpanah\thanks{farrokh@mie.utoronto.ca}}
\author{Javad Mostaghimi}
\author{Markus Bussmann}
\affil{\normalsize Department of Mechanical \& Industrial Engineering, University of Toronto, Canada}
\date{}

Published in: \href{https://doi.org/10.1080/10407790.2021.1929295}{Numerical Heat Transfer, Part B: Fundamentals, 79(5-6), pp. 255-277}

DOI: \href{https://doi.org/10.1080/10407790.2021.1929295}{10.1080/10407790.2021.1929295}

Direct Free E-Print Access: \href{https://www.tandfonline.com/eprint/74UPANRYIMDVHZGZFKFE/full?target=10.1080/10407790.2021.1929295}{Full Text}

\maketitle

\begin{abstract}
\noindent
A three-dimensional model is presented for prediction of solidification behavior using a non-linear transformation of the enthalpy equation in a Smoothed Particle Hydrodynamics (SPH) discretization. The effect of phase change in the form of release and absorption of latent heat is implemented implicitly as variable source terms in the enthalpy calculation. The developed model is validated against various experimental, analytical, and numerical results from the literature. Results confirm accuracy and robustness of the new procedure. Finally, the SPH model is applied to a study of Suspension Plasma Spraying (SPS) by predicting the impact and solidification behavior of molten ceramic droplets on a substrate.


\end{abstract}



\nomenclature{$T^*$}{Kirchhoff temperature}
\nomenclature{$T$}{Temperature}
\nomenclature{$T_m$}{Melting temperature}
\nomenclature{$T_1$}{Alloy lower melting temperature}
\nomenclature{$T_2$}{Alloy upper melting temperature}
\nomenclature{$t$}{Time}
\nomenclature{$l$}{Liquid state (subscript)}
\nomenclature{$s$}{Solid state (subscript)}
\nomenclature{$m$}{Mushy state (subscript)}
\nomenclature{$C$}{Heat capacity}
\nomenclature{$k$}{Thermal conductivity}
\nomenclature{$H$}{Enthalpy}
\nomenclature{$L$}{Latent heat}
\begin{multicols}{2}
\printnomenclature
\end{multicols}

\section{Introduction}


Solid/liquid phase transformation is an important part of many engineering applications, e.g., additive manufacturing/3D printing, surface coating, and metal casting. For example, the design of automotive and aerospace cast molds used for producing complex parts heavily relies on accurate knowledge of how fast the molten metal flowing in the narrow channels will solidify; and heat transfer and solidification rates during additive manufacturing methods such as laser powder bed fusion (LPBF) determine the quality of finished parts, thermal stresses, and deformations. An accurate knowledge of phase change process would assist in improving the quality of finished manufactured parts, while speeding the manufacturing process, leading to higher quality and more cost effective products.

Being able to model transient phase change is key to the development and optimization of these methods. While many aspects of the phase change process can be measured in the lab, prediction of the process outcome when it comes to complex geometries under different boundary conditions can be challenging. The design cost for molds and parts can be significantly reduced by limiting prototyping iterations utilizing numerical predictions.

The Finite Volume and Finite Element methods \cite{Dalhuijsen_86} are most commonly used for the numerical study of phase change. But as Smoothed Particle Hydrodynamics (SPH) offers versatility through high computation efficiency, easy parallelization, and the ability to handle complex geometries, having a robust phase change methodology would expand its current applications. SPH has been sucessfully used for applications such as the simulation of tsunamis \cite{Liu_08}, floating bodies like ships \cite{Cartwright_04}, the atomization of liquid jets \cite{Farrokhpanah_14}, the impact of liquid drops over surfaces \cite{Farrokhpanah_12_2}, and other studies of multiphase flow \cite{Hu_06,Grenier_08,Tartakovsky_09}. The following work is focused on expanding this versatility, by presenting a phase change implementation in SPH.

The amount of heat needed for the phase change process is called the latent heat. This heat is released (absorbed) during solidification (melting) at a constant temperature for pure materials, and over a temperature range for alloys. Including phase change into a numerical model requires modifying the heat transfer calculation either by adding the heat as a source term, as in Passandideh \cite{Pasandideh_96_2} and Voller \cite{Voller_87}, or by modifying the heat capacity coefficient, as in Thomas et al. \cite{Thomas_84}, Hsiao \cite{Hsiao_86}, and Dalhuijsen et al. \cite{Dalhuijsen_86}. These studies use Finite Volume and Finite Element models. The inclusion of latent heat in a SPH formulation, by modifying the heat capacity coefficient, was presented in \cite{Farrokhpanah_16_3}, but the addition of a source term implementation in an energy formulation of SPH has so far not been presented. Common practice in previous studies is the explicit inclusion of latent heat as a post-processing step after solving the energy equation, as in the works of Cleary et al. \cite{Cleary_98, Cleary_06}. This model has been applied to studies of casting \cite{Cleary_10} and droplet solidification \cite{Zhang_07,Zhang_08,Zhang_09}. As an alternative, Monaghan et al. \cite{Monaghan_05} suggested modeling the solidification of pure and binary alloys using a stationary grid of ghost particles. The drawback of this procedure is the need for extra particles, resulting in implementation difficulties.

Here, a new procedure is introduced for incorporating the latent heat into a SPH discretization, by non-linear transformation of the enthalpy equation for handling transient convective phase change problems, inspired by the work of Cao et al. \cite{Cao_89}. The latent heat here will appear implicitly as a source term in the energy calculation, and, so can be reliably applied to free surface convective problems. As will be shown, validation of the resulting formulation coupled to various analytical and numerical results prove the robustness of the procedure. The new model is then applied to a study of the impact and solidification of sub-micron ceramic droplets generated during Suspension Plasma Spraying.


\section{Governing Equations}
\subsection{Lagrangian Navier-Stokes equations}

The Navier-Stokes equations in a Lagrangian framework are
\begin{equation}\frac{D\rho}{Dt}=-\rho \nabla \cdot v\end{equation}
\begin{equation}\frac{D\mathbf{v}}{Dt} = \frac{1}{\rho}[-\nabla p+ \mu \nabla ^2 \mathbf{v}+\mathbf{F^{st}}+\mathbf{F^b} ]\end{equation}
where $\mathbf{F^b}$ represents external body forces such as gravity and $\mathbf{F^{st}}$ is the surface tension force. These equations are closed by an equation of state, which calculates pressure as a function of density~\cite{Monaghan_12}
\begin{equation}P=P_0 (\frac{\rho}{\rho_0})^\gamma+b\end{equation}
where $\gamma= 7$ and $1.4$ for liquid and gas phases respectively, $b$ is a background pressure, and $P_0$ represents a reference pressure adjusted to limit maximum density deviations from $\rho_0$ to on the order of 1\%~\cite{Monaghan_12}.

In SPH, the domain is discretized using Lagrangian particles, with each particle representing a finite fluid element. Applying integral interpolation to the continuity and Navier-Stokes equations, yields their particle form as

\begin{equation}\label{eq:H_1_p}
\frac{D\rho_i}{Dt} \cong \sum_{j=1}^{N}m_j(\mathbf{v_i-v_j}) \cdot \nabla_iW
\end{equation}

\begin{equation}
\frac{D\mathbf{v_i}}{Dt} \cong-\sum_{j=1}^{N}\left[m_j(\frac{p_i}{\rho_i^2}+\frac{p_j}{\rho_j^2})\mathbf{\nabla_i}W+ 4 m_j \left( \frac{ (\mu_i+\mu_j) \mathbf{x_{ij}} \cdot \mathbf{\nabla_i}W}{(\rho_i+\rho_j)^2 |\mathbf{x_{ij}}|^2} \right) \mathbf{v_{ij}}+\frac{\mathbf{F_{ij}^{st}}}{m_i} +\mathbf{F_i^b} \right]
\end{equation}
where $i$ denotes each particle in the domain, $j$ all particles in the neighbourhood of particle $i$, and $m_i$ is the mass of particle $i$. To improve numerical stability, the position vector that appears in both the numerator and denominator of the term $\frac{\mathbf{x_{ij}} \cdot \mathbf{\nabla_i}W}{|\mathbf{x_{ij}}|^2}$ needs to be removed to avoid division by small values.

Surface tension is calculated by applying interaction forces between all particles $i$ and $j$~\cite{Tartakovsky_05}
\begin{equation}
\mathbf{F_{ij}} = \left\{
\begin{array}{ll}
         S_{ij}cos(\frac{1.5\pi}{3h}|\mathbf{x_{ij}}|)\frac{\mathbf{x_{ij}}}{|\mathbf{x_{ij}}|} & |\mathbf{x_{ij}}|\leq h \\
        0 & |\mathbf{x_{ij}}| >  h

        \end{array} \right.
\end{equation}
$S_{ij}$ is a constant that controls the magnitude of the surface tension force between each phase. Contact angles, and the interaction between different fluids and solids, can be imposed by specifying different values of $S_{ij}$.

For particles that come in contact with a wall, the no-slip boundary condition must be applied. The method of Holmes et al. \cite{Holmes_11} has been implemented here, as in most cases studied later, the fluid comes in contact with a wall at relatively large velocity.

Finally, SPH particles subjected to negative pressure tend to form small clusters. While most simulation results are not significantly affected by this rearrangement, it can become important for the test cases of interest here. Large particle impact velocities lead to large displacements in particle positions, and as a solidification front moves across a particle, the freezing of some particles in a cluster can lead to unstable solutions. The convection term of Monaghan \cite{Monaghan_00} is added to the momentum equation, to adjust particles when negative pressures are encountered.

\subsection{Energy equation}

The energy equation is solved using the methodology proposed by Cao et al. \cite{Cao_89}. The three-dimensional energy equation for laminar flow with no viscous dissipation is

\begin{equation}\label{eq:H_4}
\frac{\partial}{\partial t}\left( \rho H \right)+\nabla \cdot \left( \rho \mathbf{v} H \right)=\nabla \cdot \left( k \nabla T \right)
\end{equation}

and for an incompressible fluid,
\begin{equation}\label{eq:H_5}
\frac{dH}{dT}=C(T)
\end{equation}

Enthalpy is calculated with respect to a reference point, which we choose to be the melting temperature of the solid material, at which the enthalpy is assumed to be zero (see figure \ref{fig:H_1}).

\subsubsection{Temperature Calculation}

From equation \ref{eq:H_5} and the choice of a reference point, temperature can be calculated. For a pure substance with phase change at a single temperature

\begin{equation}\label{eq:H_6}
T = \left\{ \begin{array}{lll}
        T_m+ H/C_s & \mbox{$H \leq 0$}\\
        T_m & \mbox{$0 < H < L$}\\
        T_m + (H-L)/C_l & \mbox{$H \geq L$}\\
\end{array} \right.
\end{equation}
and for materials like alloys where phase change occurs over a temperature range

\begin{equation}\label{eq:H_7}
T-T_1 = \left\{ \begin{array}{lll}
        H/C_s & \mbox{$H \leq 0$} & \mbox{$solid$}\\
        \Delta T H/(L+C_m \Delta T) & \mbox{$0 < H < L+C_m \Delta T$} & \mbox{$mushy$}\\
       H/C_l - \left[ L + \left( C_m - C_l \right) \Delta T \right] / C_l & \mbox{$H \geq L+C_m \Delta T$} & \mbox{$liquid$}\\
\end{array} \right.
\end{equation}

As shown in figure \ref{fig:H_1}, for pure materials, enthalpy experiences a step jump at the melting temperature $T_m$. For alloys with a phase change temperature interval (PCTI), phase change occurs over a temperature range $T_1$ to $ T_2$. When the temperature is within the PCTI, the material is mushy. In equation \ref{eq:H_7}, enthalpy is assumed to vary linearly over the PCTI.

\subsubsection{Kirchhoff temperature}

To eliminate the constant values in equations \ref{eq:H_6} and \ref{eq:H_7}, the Kirchhoff temperature is defined as $T^*=\int_{T_{m}}^{T} k(\zeta ) d \zeta$. Applying $T^*$ to equations \ref{eq:H_6} and \ref{eq:H_7} gives

\begin{equation}\label{eq:H_8}
T^* = \left\{ \begin{array}{lll}
       k_s (T-T_m) & \mbox{$T < T_m$}\\
        0 & \mbox{$T=T_m$}\\
     k_l (T-T_m) & \mbox{$T>T_m$}\\
\end{array} \right.
\end{equation}
for a pure material and

\begin{equation}\label{eq:H_9}
T^* = \left\{ \begin{array}{lll}
       k_s (T-T_1) & \mbox{$T \leq T_1$}\\
        k_m ( T- T_1) & \mbox{$T_1 < T <T_2$}\\
     k_l (T-T_1) & \mbox{$T \geq T_2$}\\
\end{array} \right.
\end{equation}
for a material with a PCTI.

The constant values $T_m$ and $T_1$ can now be eliminated by substituting $T$ from equations \ref{eq:H_6} and \ref{eq:H_7} into equations \ref{eq:H_8} and \ref{eq:H_9}, respectively. This gives the Kirchhoff temperature as a function of enthalpy in the form

\begin{equation}\label{eq:H_10}
T^* = \left\{ \begin{array}{lll}
       k_s H / C_s & \mbox{$H \leq 0$}\\
        0 & \mbox{$0 < H < L$}\\
     k_l (H-L)/C_l & \mbox{$H \geq L$}\\
\end{array} \right.
\end{equation}
for a pure material and

\begin{equation}\label{eq:H_11}
T^* = \left\{ \begin{array}{lll}
       k_s H / C_s & \mbox{$H \leq 0$}\\
        k_m \left[ \Delta T H / (L+C_m \Delta T) \right] & \mbox{$0 < H < L+C_m \Delta T$}\\
     k_l \left[ H/C_l - \left[ L + (C_m - C_l) \Delta T \right] /C_l \right] & \mbox{$H \geq L + C_m \Delta T$}\\
\end{array} \right.
\end{equation}
for a material with a PCTI. The linearity of equations \ref{eq:H_10} and \ref{eq:H_11} allows the easy calculation of $H$ or $T^*$ when only one is known. Expressing $T^*=\Gamma (H)H+S(H)$, $\Gamma$ and $S$ can be derived from equations \ref{eq:H_10} and \ref{eq:H_11} as

\begin{equation}\label{eq:H_12}
\begin{array}{lll}
\Gamma(H) = \left\{ \begin{array}{lll}
       k_s / C_s & \mbox{$H \leq 0$}\\
        0 & \mbox{$0 < H < L$}\\
     k_l /C_l & \mbox{$H \geq L$}\\
\end{array} \right.
&
S(H) = \left\{ \begin{array}{lll}
       0 & \mbox{$H \leq 0$}\\
        0 & \mbox{$0 < H < L$}\\
     -L k_l / C_l & \mbox{$H \geq L$}\\
\end{array} \right.
\end{array}
\end{equation}
for single temperature phase change and

\begin{equation}\label{eq:H_13}
\begin{array}{lll}
\Gamma(H)= \left\{ \begin{array}{lll}
       k_s / C_s & \mbox{$H \leq 0$}\\
        k_m \Delta T  / (L+C_m \Delta T) & \mbox{$0 < H < L+C_m \Delta T$}\\
     k_l  /C_l & \mbox{$H \geq L + C_m \Delta T$}\\
\end{array} \right.
\\\\
S(H) = \left\{ \begin{array}{lll}
       0 & \mbox{$H \leq 0$}\\
       0 & \mbox{$0 < H < L+C_m \Delta T$}\\
     k_l \left[ L + (C_m - C_l) \Delta T \right] /C_l & \mbox{$H \geq L + C_m \Delta T$}\\
\end{array} \right.
\end{array}
\end{equation}
for phase change over a temperature range.

\subsubsection{Reconstruction of the Lagrangian energy equation}

With the definition of $T^*$, equation \ref{eq:H_4} can be rewritten as




\begin{equation}\label{eq:H_16}
\frac{\partial}{\partial t}(\rho H) + \boldsymbol{\nabla} \cdot (\rho H \mathbf{v} ) = \nabla^2 T^*
\end{equation}

By expanding $\boldsymbol{\nabla} \cdot (\rho H \mathbf{v} )$ into

\begin{equation}\label{eq:H_17}
\frac{\partial}{\partial t}(\rho H) + \mathbf{v} \cdot  \boldsymbol{\nabla} (\rho H)+ (\rho H)  \boldsymbol{\nabla} \cdot \mathbf{v}= \nabla^2 T^*
\end{equation}
the first two terms on the left hand side can be combined to form a total derivative

\begin{equation}\label{eq:H_18}
\frac{D (\rho H)}{Dt}=- (\rho H)  \boldsymbol{\nabla} \cdot \mathbf{v}+ \nabla^2 T^*
\end{equation}

Equation \ref{eq:H_18} is the Lagrangian form of the transformed enthalpy equation, which can be discretized to give the variation of $\rho H$ with time for each SPH particle. Finally, introducing the Lagrangian continuity equation \ref{eq:H_1_p}, the $- \rho  \boldsymbol{\nabla} \cdot  \mathbf{v}$ on the right hand side of equation \ref{eq:H_18} can be replaced by the total derivative of density, resulting in

\begin{equation}\label{eq:H_19}
\frac{D (\rho H)}{Dt}= H \frac{D \rho}{Dt} + \nabla^2 T^*
\end{equation}

\subsection{SPH discretization}

The enthalpy equation \ref{eq:H_19} accounts for the effects of phase change latent heat, and is now discretized using an SPH formulation in the form of

\begin{equation}\label{eq:H_30}
\frac{D(\rho H)_i}{Dt}=H_i \frac{D\rho_i}{Dt}+(\nabla^2T^*)_i=H_i \frac{D\rho_i}{Dt}+(\nabla^2 \Gamma H)_i + (\nabla^2 S)_i
\end{equation}
where the values for $H_i$ and $D\rho_i/Dt$ are known for each particle $i$. The Laplacians of $\Gamma H$ and $S$ must be discretized and calculated in SPH form. Following classic SPH formulations, these Laplacians can be expressed as

\begin{equation}\label{eq:H_31}
(\nabla^2 \Gamma H)_i + (\nabla^2 S)_i = \boldsymbol{\nabla_i} \cdot ( \boldsymbol{\nabla_i} (\Gamma H) + \boldsymbol{\nabla_i} S )
\end{equation}
The inner gradient can be calculated using

\begin{equation}\label{eq:H_29}
\boldsymbol{\nabla}f(x_i)=   \sum_{j}^{N} \frac{m_j}{\rho_i}   \left[ f(x_j)-f(x_i) \right] \cdot  \boldsymbol{\nabla_i}W_{ij}
\end{equation}

while the outer divergence is calculated as

\begin{equation}\label{eq:H_28}
\boldsymbol{\nabla \cdot \Phi(x_i)}=\rho_i \left[  \sum_{j}^{N} m_j    \left( \frac{\boldsymbol{\Phi(x_i)}}{\rho_i^2} +\frac{\boldsymbol{\Phi(x_j)}}{\rho_j^2} \right) \cdot  \boldsymbol{\nabla_i}W_{ij}  \right]
\end{equation}
As a result, the inner gradient in equation \ref{eq:H_30} becomes

\begin{equation}\label{eq:H_32}
\boldsymbol{G_i}\equiv\boldsymbol{\nabla_i}(\Gamma H) + \boldsymbol{\nabla_i}S=\sum_{j}^{N} \frac{m_j}{\rho_i}
\left\{
\left[
\Gamma(H_j)H_j + S(H_j)
\right]-\left[
\Gamma(H_i)H_i + S(H_i)
\right]
\right\} \boldsymbol{\nabla_i W_{ij}}
\end{equation}

Here, values of $\Gamma$ and $S$ are calculated using equation \ref{eq:H_12} or \ref{eq:H_13}, using values of $H$ for the particles $i$ and $j$. Then the outer divergence in equation \ref{eq:H_31} becomes

\begin{equation}\label{eq:H_33}
(\boldsymbol{\nabla \cdot G})_i=\rho_i \left[
\sum^{N}_{j} m_j
\left(
\frac{\boldsymbol{G_i}}{\rho_i^2} + \frac{\boldsymbol{G_j}}{\rho_j^2}
\right) \cdot \boldsymbol{\nabla_i W_{ij}}
\right]
\end{equation}

This classic SPH discretization may become unstable when there is a discontinuity in the physical properties in the domain \cite{Cleary_99}, as for example during heat transfer between two materials with different thermal conductivities. To overcome this issue, the remedy of Cleary et al. \cite{Cleary_99} is extended and applied here. To begin, the Laplacians in equation \ref{eq:H_30} are directly calculated using

\begin{equation}\label{eq:H_33_1}
\nabla^2 (\Gamma H + S)_i= \sum_{j}^{N} 2 \frac{m_j}{\rho_j}  \left(  f_i - f_j   \right) 	\frac{\boldsymbol{\nabla_i}W_{ij}}{\boldsymbol{r_{ij}}}
\end{equation}
where the variable $f$ is defined as

\begin{equation}\label{eq:H_33_2}
\begin{array}{ll}
f_i = \Gamma_{eff}  H_i  +  S_i\\
f_j = \Gamma_{eff}  H_j  +  S_j\\
\end{array}
\end{equation}
and $\Gamma_{eff}$ is defined as

\begin{equation}\label{eq:H_33_3}
\Gamma_{eff}  = \left\{ \begin{array}{ll}
       \frac{2 \Gamma_i \Gamma_j}{\Gamma_i + \Gamma_j} & \mbox{$\Gamma \neq 0$}\\
     0 & \mbox{$otherwise$}\\
\end{array} \right.
\end{equation}
Substituting back into equation \ref{eq:H_30}, the variation of $\rho H$ for particle $i$ is calculated as

\begin{equation}\label{eq:H_34}
\frac{D(\rho H)_i}{Dt} = H_i \left( \frac{D \rho}{Dt} \right)_i + \nabla^2 (\Gamma H + S)_i
\end{equation}

Unless otherwise stated, the smoothing kernel proposed by Meng et al. \cite{Meng_14} presented in table \ref{tab:H_1} is used throughout this paper.

\subsection{Mushy region treatment}

As mentioned, the melting and solidification of an alloy occurs through a mushy zone due to the difference in phase change temperatures of elements forming the alloy. One element will start to solidify first, in the form of solid crystals that nucleate and grow into a slurry of solid and molten material. This phenomenon can happen in different ways. In one, the solidified phases might grow at discrete sites, leading to a suspension of solid phases in a liquid phase. But solidification might also occur as a solid front growing from a wall, with small scale porous branches that spread into the still molten phase. In this case, the liquid phase will flow through the porous solid phase that is growing in size, and be similar to the flow of a liquid phase through a porous material. A more detailed discusssion of these conditions can be found in Voller et al. \cite{Voller_90}.

The effect of a mushy region is typically included in one of two ways. When the flow is similar to flow through a porous media, a Darcy source term can be added to the momentum equation, to account for the pressure drop due to fluid moving through pores. This flow is often found in casting problems \cite{Voller_90}. When the flow is more homogeneous, the effect of the slurry-like mixture is accounted for by modifying the viscosity. Depending on the value of temperature or enthalpy, a liquid fraction $\alpha$ and a solid fraction $\beta = 1 - \alpha$ can be defined for each particle at each iteration. Based on these values, the viscosity of each particle is updated using \cite{Voller_90}
\begin{equation}\label{eq:H_35}
\mu = \alpha~\mu_{liquid} + \beta~\mu_{solid}
\end{equation}
For particles that have a liquid fraction of unity, the viscosity will be equal to the viscosity of the molten material. For a liquid fraction of zero, the viscosity of each particle is set to a large value to mimic solid matter. In between these two limits, the viscosity varies linearly. The choice of a linear variation is in contrast to more complex variations; for example \cite{alavi_12}
\begin{equation}\label{eq:H_36}
\mu = \mu_{liquid} \left(     1 + C  \frac{\beta^3}{(1-\beta)^3 + \epsilon}     \right)
\end{equation}
where $C$ is a problem-dependent constant and $\epsilon$ is introduced to avoid division by zero. As will be shown, equation \ref{eq:H_35} yields more stable solutions in the framework presented here, and hence has been utilized in this study.

\section{Validation and Results}

\subsection{1D Conduction}

Similar to Cleary et al. \cite{Cleary_99}, consider an infinite slab with a temperature discontinuity at $x=0$. The material properties are $\rho_1$, $k_1$, $C_1$ when $x<0$ and $\rho_2$, $k_2$, $C_2$ when $x>0$. With an initial temperature jump of $T_{01}$ across the discontinuity, the transient analytical solution to this problem is \cite{Carslaw_59}

\begin{equation}\label{eq:H_analytical_1}
T-T_{01}= \left\{ \begin{array}{ll}
        \frac{k_1 \alpha_1^{-1/2} T_0}{k_1 \alpha_1^{-1/2}+k_2 \alpha_2^{-1/2}}  \left[     1 + \frac{k_2 \alpha_2^{-1/2}}{k_1 \alpha_1^{-1/2}}  {erf}\left( \frac{x}{2 \sqrt{\alpha_1 t}} \right)      \right]        & \mbox{$x>0$}\\
        \frac{k_1 \alpha_1^{-1/2} T_0}{k_1 \alpha_1^{-1/2}+k_2 \alpha_2^{-1/2}}  \left[       {erfc}\left( \frac{|x|}{2\sqrt{\alpha_2 t}}  \right)      \right]        & \mbox{$x<0$}\\
\end{array} \right.
\end{equation}

Here, $\alpha = k / \rho C$ is the thermal diffusivity. To avoid the complications of modeling an infinite slab, a slightly modified problem is modeled. The $x$ domain is limited to $-1< x <1$ and a no-heat flux boundary condition is applied at each end. It is evident that the temperature profile will be slightly different than the analytical solution near the $x$ limits, especially as time goes on, but the temperature throughout the domain should converge to $T_0/2$. Two test cases were run with $80$ particles in the $x$ direction. For the first case, the temperature jump and all thermal properties are set to unity. Figure \ref{fig:H_2_1} shows the temperature profile for this case at various times. The solid lines in this figure are the analytical solutions obtained from equation \ref{eq:H_analytical_1}. As expected, temperature values near $-1$ and $1$ start to deviate from the analytical solution as time progresses. The values of temperature in the middle of the domain, however, are a good match to the analytical solution.

The second case includes a discontinuity in the thermal properties. The initial temperature jump is set to 10, and the thermal properties are the same as before, except $k_1=10$. Figure \ref{fig:H_2_2} clearly shows that at small times the simulation exactly predicts the analytical solution, but as time goes on, the error increases near the boundaries. Error values for this test case, calculated as $L_2 = \sqrt{ \frac{1}{N}  \sum_{i=1}^N {     \left(   T_i^{SPH} -  T_i^{exact}   \right)^2     }  }$, are plotted in figure \ref{fig:H_2_3}.

\subsection{2D Conduction}
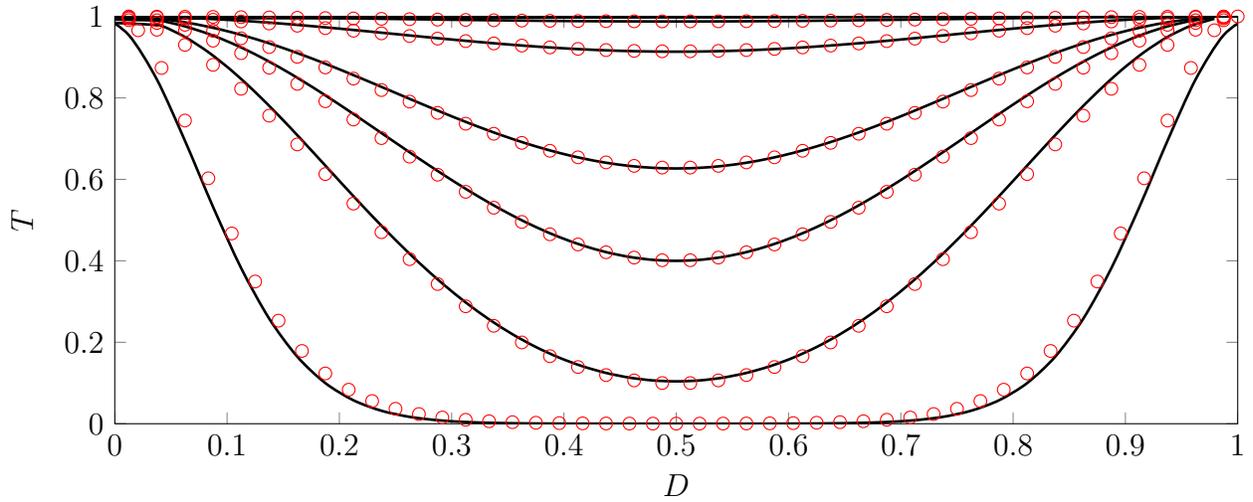
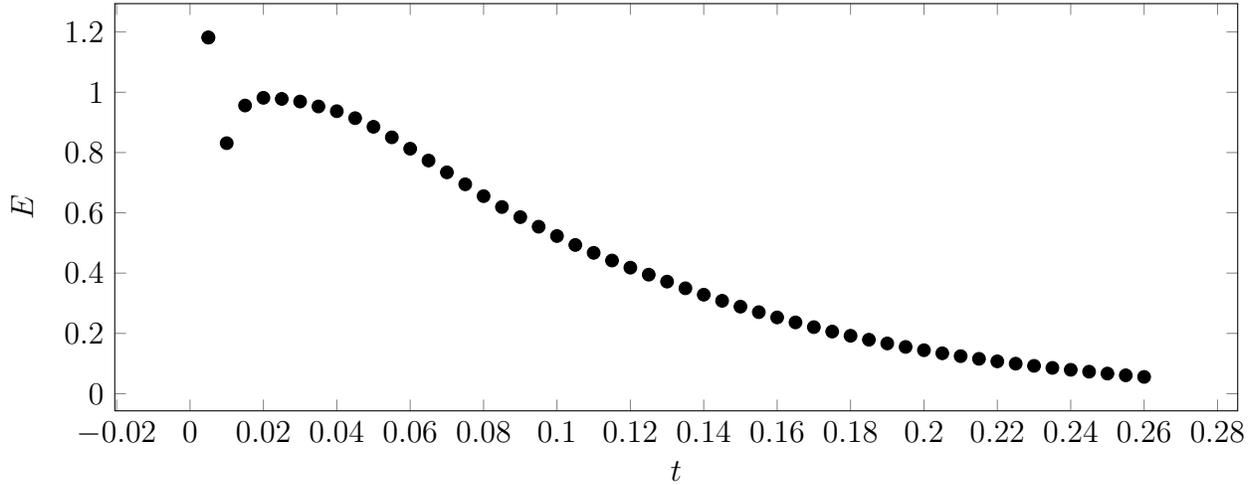
In this case, a two-dimensional square plate of side length 1 is considered. The initial temperature is $0$, at $t=0$ the temperature on the four sides is set to 1. For simplicity, the density, thermal conductivity, and heat capacity are all set to unity. The results of the SPH solver are compared to a result generated by Ansys Fluent. The transient temperature profile on the diagonal of the square, and the error values, are plotted in figures \ref{fig:H_2_4} and  \ref{fig:H_2_5}, respectively. Results show excellent agreement between the SPH and Fluent solvers.

\subsection{2D Phase Change}

We now consider a 2D problem that involves phase change. For the test cases described below, SPH particles are uniformly distributed with a spacing $\Delta x=\Delta y$ over the cross-section of a square, as shown in figure \ref{fig:H_23}. The domain is periodic in the $z$ direction. Since each particle has a neighborhood with a radius of $3 \Delta x$, 6 layers of particles are placed on the wall, so that all inner particles have a complete neighborhood. The outer 3 layers are added to assure valid gradients are calculated at the 3 inner layers of the wall. The temperature of all wall particles is kept constant throughout all iterations.

\subsubsection{Comparison to numerical results}

Numerical results available in the literature for the so-called Stefan problem are used as validation of the SPH results including phase change. To simulate this problem, the domain is initially filled with liquid; solidification starts when the temperature of the walls drops below freezing. All properties for solid and liquid are assumed to remain constant in each phase.



Values of the non-dimensional parameters are $\theta \equiv (T_i-T_w)/(T_m-T_w)=1$ and $St \equiv C_s (T_m-T_w)/L=0.641$. Figure \ref{fig:H_4} shows the solidification front along the diagonal of the domain. For comparison, results reported by Cao et al. \cite{Cao_89}, Crowley \cite{Crowley_78}, and Hsiao et al. \cite{Hsiao_86} are also plotted in this figure. Results show the trends reported by the SPH model agree well with the literature.

The second problem is the same, but the initial temperature is higher than the phase change temperature. Non-dimensional parameters for this problem are $\Theta=9/7$ and $St=2$, $\alpha _l/\alpha _s=0.9$, and $k_l/k_s=0.9$. Figure \ref{fig:H_5} compares results for this test case against results of Cao et al. \cite{Cao_89}, Hsiao et al. \cite{Hsiao_86}, and Keung \cite{Keung_80}. SPH values are again in excellent agreement with the literature values.

\subsubsection{Effect of particle resolution}

To make sure the solution is not dependent on particle resolution, the above problem were run at different resolutions. Figure \ref{fig:H_6} shows results as the resolution is changed from $10\times 10$ to $100\times 100$. It can be seen that by increasing the resolution, results converge to unique values, and then results remain nearly identical at resolutions above these values.

\subsubsection{Effect of $\Delta T$}

The same test case was calculated using the same properties, but for phase change that occurs over a temperature range $\Delta T=20$. The particle resolution is $100\times 100$. The results shown in figure \ref{fig:H_5} are not sensitive to allowing the phase change to occur over a temperature range. The solidification front here is assumed to be at the liquid fraction of 0.5.

\subsubsection{Effect of Smoothing Kernel}

Finally, the effect of different smoothing functions was studied, considering the common functions shown in table \ref{tab:H_1}. The first example from before was re-run using each of these kernels. Figure \ref{fig:H_7} shows that the choice of smoothing kernel does not affect the results.

\subsection{3D Problem: Droplet Impact and Solidification}

The three-dimensional problem here is the impact and solidification of a molten drop, similar to the conditions of Aziz et al. \cite{Aziz_00} and Passandideh et al. \cite{Pasandideh_02}. In this problem, a molten tin drop impacts onto a stainless steel substrate. The drop is initially at \ang{240}C. The substrate is at room temperature. The impact velocity is 1m/s, and the drop has an initial diameter of $2.7mm$. The spread factor ($D/D_0$) for this test case is plotted in figure \ref{fig:H_8}. The experimental results of Aziz et al. \cite{Aziz_00} and the numerical result of Pasandideh et al. \cite{Pasandideh_02} are plotted for comparison. Results indicate a good agreement to the experimental and numerical results for spread factor of a low velocity impacting drop. The final SPH prediction of the spread factor is closer to the numerical results of Passandideh et al., both deviating about 6\% from the experimental value.

\subsection{Mushy zone solidification}

Identifying a benchmark for the mushy zone implementation was not possible. There are no robust and accurate experimental measurements of a mushy region. As for numerical studies of mushy zone solidification, most studies are of natural convection, where a material solidifies in a cube (or square in 2D). During solidification, the natural convection resulting from temperature variations accelerates molten liquid to flow in a manner similar to flow in a lid-driven cavity. This results in an uneven growth of the solid front, that can be used as a means of evaluating predictions of mushy zone solidification. Figure \ref{fig:mushy_1} shows a schematic of this problem. The liquid phase is confined within the cavity. The lower and upper walls are adiabatic. The fluid exchanges heat with only the cold and hot opposite walls. To speed computation time for the flow pattern to form, a divergence free velocity field $v_{\theta}=20 \frac{r}{R}$ and $v_r=0$ is imposed at $t=0$. Figure \ref{fig:mushy_1} illustrates the initial velocity field in the domain.

The solidification fronts at $t=0.16$, corresponding to $\alpha = 0.5$ for viscosity predictions using equations \ref{eq:H_35} and \ref{eq:H_36}, are plotted in figure \ref{fig:mushy_3}. Particle positions at $t=0.16$, after solidification has been partially completed, are shown in figure \ref{fig:mushy_2}. Although the formulations yield similar results, the particle spacings and arrangements are more realistic using \ref{eq:H_35}. Hence, equation \ref{eq:H_35} was used to capture mushy effects in all subsequent results.
\subsection{Application to Suspension Plasma Spraying (SPS)}

The developed SPH model is applied here to the study of surface coating by impact and solidification of droplets generated by the Suspension Plasma Spraying (SPS) process. SPS is used for depositing high-quality thermal barrier coatings. A fine ceramic powder is suspended in a liquid such as water or ethanol. Injecting this suspension into a plasma flow, the heat of the plasma evaporates the carrier fluid and then melts the solid particles, which then impact a substrate at high velocity and solidify. SPS is of particular interest because of the small size and high speed of the droplets that impact a substrate. The finished solidified surface is affected by the balance between the speed of droplet spread on the surface and the solidification speed. Slow solidification will allow droplets to spread or recoil freely on a surface, as opposed to fast solidification.

The SPH model was used to study and compare the time-scales at which these high speed impacts and solidifications occur, based on \cite{Farrokhpanah_16_2} that contains size, velocity, temperature, and position data of a million SPS droplets generated from a Yttria-Stabilized Zirconia (YSZ) suspension in an argon plasma flow. YSZ is a ceramic with high melting temperature suitable for producing advanced thermal barrier coatings. In what follows, we present SPH results of the impact of one, two, and five SPS droplets onto a surface to learn how they spread, solidify, and interact.




\subsubsection{Spread Factor}

To predict the final conditions of a coating, the SPH solver was used to calculate spread factor at $50$, $100$, and $200m/s$, which covers the velocity range of YSZ droplets impacting a substrate during SPS. Results are plotted in figure \ref{fig:SPS_IMP_11} as a function of non-dimensional time ($t^*=tV/D_0$). These velocities correspond to non-dimensional values of 96, 192, and 383 for $Re=\frac{\rho V_0 D_0}{\mu}$, and 323, 1293, 5172 for $We=\frac{\rho V_0^2 D_0}{\sigma}$. Since the Weber numbers are much larger than the Reynolds numbers, i.e. $We \gg \sqrt{Re}$, capillary effects can be neglected \cite{Pasandideh_96}, and as Passandideh-Fard et al. \cite{Pasandideh_96} show for $We \gg 12$, maximum spread factor can be approximated as $0.5 Re^{0.25}$. Results obtained here suggest a similar correlation for maximum spread factor $\xi_{max} = \frac{D_{max}}{D_0} = 0.59 ~ Re ^{0.28}$ ($R^2=0.9960$)

\subsubsection{Solidification Time}
Here we consider the solidification time and amount of heat transferred from a molten YSZ droplet to a substrate. A $3 \mu m$ diameter droplet impacts a substrate at $150 m/s$. The molten droplet temperature is $3000K$ and the substrate is at $600 K$. Two different substrate materials are considered. The first is YSZ, identical to the droplet material, with a thermal conductivity of $2.4 W/(m \cdot K)$. For the second, the thermal conductivity of the substrate set to $66 W/(m \cdot K)$, similar to tin, mimicking the formation of an initial coating layer, when droplets first impact a metallic substrate. Figure \ref{fig:SPS_IMP_11_25} shows the impact and solidification of the YSZ substrate. Solidification begins at the bottom of the splat, and grows faster near the splat edges than in the middle. The bulk of the splat solidifies last as the heat in the bulk faces a larger thermal resistance as it escapes into the substrate. Finally, the low thermal conductivity of YSZ means that it takes a long time to reach a homogeneous temperature distribution within the splat, which is only fully solidified by $t^*=1500$.

The heat flux into a metallic substrate is also of particular interest. The low thermal conductivity of YSZ slows heat transfer from splats to the substrate during spraying. A high thermally conductive substrate promotes the initial heat transfer as the metallic piece with a high Biot number is capable of maintaining a more uniform temperature distribution. This keeps the surface temperature closer to the temperature of the metallic substrate, and therefore, enhances the heat flux from the splat into the substrate due to a higher temperature difference between the interfaces. Figure \ref{fig:SPS_IMP_11_5} shows the heat flux at the impact point of a droplet onto a substrate with the higher thermal conductivity, as a function of non-dimensional time. Results show that upon impact, the heat transfer to the substrate jumps to a very large value, because the substrate and droplet are at the highest temperature difference, and the impact area between the two is minimum at impact. As the droplet spreads over the surface, the area grows and distributes heat, and so the temperature difference between the substrate and droplet decreases, causing a decrease in heat flux. The high thermal conductivity of the substrate here contributes to the rapid formation of a solidified layer next to the boundary, but the material in the bulk of each splat cools slowly, due to the low thermal conductivity of the YSZ.

\subsubsection{Void Formation}
Figure \ref{fig:SPS_IMP_13_main} illustrates results of the setup used to study cases where droplet impacts are prone to void formation, as occurs when a droplet impacts near a previously-deposited splat. For the test cases below, a droplet of radius $3 \mu m$ impacts onto a surface with a step, of height 0, 0.12, 0.24, 0.36, and $0.54 \mu m$.

In the first case, figure \ref{fig:SPS_IMP_12}, the initial temperature of the droplet is $2985 K$, only slightly above the solidification point of YSZ and so solidification occurs immediately upon contact with the substrate. This prevents any significant pore formation near the step. However, SPS droplets are typically hotter. Hence the same test case, but with an initial temperature of $3500 K$, is shown in Figure \ref{fig:SPS_IMP_13}. In this case, the droplet takes longer to solidify, and so spread and recoil occur while the droplet is still molten. The impact this time is controlled by the fluid momentum of the drop spreading over the step, and surface tension acting upon it. The pore that forms after the spread is driven by the speed of solidification and the surface tension acting on the nearly stagnant splat at the time of solidification. Here the pore can be approximated by a quarter circle, with a radius close to the step height. It is evident that at the lower temperature, close to the melting point of YSZ, immediate solidification prevents pore formation.

\subsubsection{Impact of Two Droplets}
Results reported by \cite{Farrokhpanah_16_2} on the SPS of YSZ droplets indicate that many of the binary impacts of splats that occur on a substrate involve $4-6 \mu m$ droplets, because these are large enough to create splats that overlap. To examine the interaction of droplets, results are presented of two impacts. The first is a $4.59 \mu m$ droplet impacting a substrate at $175 m/s$ and $3517 K$. The solidified splat is shown in figure \ref{fig:SPS_IMP_11_6}. The impact and spread of this drop on the substrate is completed by $t^*=110$, but it then takes until $t^*=1500$ for the splat to fully solidify.

To examine binary interaction, a second $6.8 \mu m$ droplet impacts onto the first at a velocity of $231 m/s$ and temperature of $3795 K$, where the point of impact is $14 \mu m$ from the center of impact of the first droplet. The result of this second impact is shown in figure \ref{fig:SPS_IMP_11_7}. The first splat has solidified by the time the second droplet impacts. The second droplet spreads more in the free directions away from the first splat. Towards the first splat, the second splat spreads in three directions, two bypassing the first splat, and one that splashes over it.

\subsubsection{Multiple Impacts}
The SPS process is very efficient at coating a surface, as the mist of droplets generated from atomization of the liquid suspension uniformly covers a large area, making it very unlikely to get multiple neighbouring impacts within a short time interval. This means that there is enough time between impacts for splats to solidify before being impacted by subsequent droplets. The SPS data of \cite{Farrokhpanah_16_2} confirms this: the time between successive impacts is larger than $t^*=300,000$, while solidification takes place on the order of $t^* \sim 1000$. This assumption is taken into account for the study of multiple impacts here. From the SPS data, five droplets were chosen that overlapped. The location and properties of these droplets at the time of impact are provided in table \ref{tab:SPS_IMP_1}. As is evident, these droplets are of different size.

Note that velocity components parallel to the substrate, though small in magnitude, are included in this simulation. For the simulations here, the substrate is YSZ. Though at the start of a coating process, the substrate material will play an important role, as metallic substrates have a higher thermal conductivity than YSZ, as YSZ coats the surface over time, subsequent droplets will mostly impact a substrate covered by YSZ. The droplets on impact are assumed to be molten. Mushy zone effects are included using equation \ref{eq:H_35}.

Results of this case are presented in figures \ref{fig:SPS_IMP_0} and \ref{fig:SPS_IMP_8}. Figures are colored by the $z$ magnitude at each point, identifying how the roughness of the coating changes due to the increase in local height. The first two droplets, shown in figure \ref{fig:SPS_IMP_0}, have larger diameters of $9 \mu m$ with impact velocities higher than $200 m/s$. Hence, the splats cover a large area. The next two droplets have smaller diameters of $3 \mu m$ and lower impact velocities below $100 m/s$. The last droplet is also small, but impacts at the highest speed. The final coated substrate after five impacts is plotted in figure \ref{fig:SPS_IMP_8}. The maximum height from the substrate surface is 0.91, 1.76, 1.53, 2.40, and 2.14 $\mu m$, for droplets \#1 to \#5, respectively. Hence, the highest peak in the finished coating corresponds to a height of $2.40 \mu m$, which is achieved through overlapping splats of droplets \#1, \#2, and \#4.


\section{Conclusions}
An approach to modeling phase change using SPH has been presented. The release and absorption of latent heat during phase change is accounted for by a non-linear transformation of the enthalpy equation, by adding source terms. The model was validated against various available analytical, experimental, and numerical results from the literature, that show that the enthalpy formulation provides a reliable platform for predicting phase change. The methodology can also be applied to pure materials, without the need to define an artificial phase change interval.

The new SPH formulation was applied to a coating process known as Suspension Plasma Spraying (SPS). Combining experimental data of YSZ droplets impacting a substrate with splat formation and solidification results obtained from SPH simulations shows that the surface is coated by isolated splats. Impacted droplets impact and recoil within $t^*(=t v/D)$ of 10; cooling down and solidification take much longer, $t^* \sim 1000$, due to the high temperature of the molten ceramic at impact. SPS data indicates that the time between successive impacts on the same location of a substrate is larger than $t^*=300,000$, and so it will be rare for two molten droplets to interact on the substrate while both are still molten.

Findings also suggest a noticeable difference between the first coating layer compared to subsequent ones. For YSZ droplets that come in direct contact with a metallic substrate, the higher thermal conductivity of the substrate material will contribute to the fast formation of a solid layer beneath the splat, that will quickly immobilize it. For subsequent layers of coating, where YSZ droplets impact onto previously solidified YSZ splats, the lower thermal conductivity results in slower cooling, allowing the better spread of droplets on the substrate.

\bibliographystyle{unsrtdin}
\bibliography{References}

\clearpage
\begin{table}[!htbp]
\caption[Smoothing kernels used in various test cases]{Smoothing kernels used in various test cases}
\label{tab:H_1}
\begin{tabular*}{\textwidth}{c @{\extracolsep{\fill}} ccccc}
 \hline
\\
 \multicolumn{2}{c}{Smoothing Kernels} \\\\
 \hline
Meng et al. \cite{Meng_14} &
$
 W(R,h) =\frac{21}{16\pi h^3} \left\{ \begin{array}{ll}
( 1-R/2)^4(2R+1) & \mbox{$0 \leq R \leq 2$}\\
0 & \mbox{$otherwise$}\\
\end{array} \right.
$
\\\\
Johnson et al. \cite{Johnson_96} &
$
 W(R,h) =\frac{5}{4\pi h^3} \left\{ \begin{array}{ll}
(3/16)R^2-(3/4)R+3/4 & \mbox{$0 \leq R \leq 2$}\\
0 & \mbox{$otherwise$}\\
\end{array} \right.
$ \\\\
Monaghan et al. \cite{ Monaghan_85} &
$
 W(R,h) =\frac{3}{2\pi h^3} \left\{ \begin{array}{ll}
2/3-R^2+R^3/2 & \mbox{$0 \leq R < 1$}\\
(1/6)(2-R)^3 & \mbox{$1 \leq R < 2$}\\
0 & \mbox{$otherwise$}\\
\end{array} \right.
$  \\\\
Liu et al. \cite{Liu_03} &
$
 W(R,h) =\frac{1}{4\pi h^3} \left\{ \begin{array}{ll}
(2-R)^3-4(1-R)^3  & \mbox{$0 \leq R \leq 1$}\\
(2-R)^3  & \mbox{$1 < R \leq 2$}\\
0 & \mbox{$otherwise$}\\
\end{array} \right.
$  \\
 \hline
\end{tabular*}
\end{table}

\begin{table}[!htbp]
\caption[YSZ particle properties at impact]{YSZ particle properties at impact \cite{Farrokhpanah_16_2}}
\label{tab:SPS_IMP_1}
\begin{tabular*}{\textwidth}{@{\extracolsep{\fill}}cccccccc}
 \toprule[1mm]
~ & \multicolumn{2}{c}{Impact Position} & \multicolumn{3}{c}{Impact Velocity} & Temperature & Diameter \\
\# & $x~(\mu m)$ & $y~(\mu m)$ & $u~(m/s)$ & $v~(m/s)$ & $w~(m/s)$ & $T~(K)$ & $D~(\mu m)$ \\
 \hline
1 & -6.6 & 3.0 & -4.1 & 1.5 & -207.6 & 3679.8 & 9.1\\
2 & 1.4 & -2.9 & -0.2 & 5.9 & -226.1 & 3833.5 & 8.7\\
3 & 10.8 & -9.3 & 27.7 & 5.1 & -88.5 & 3523.5 & 3.0\\
4 & 4.7 & 5.7 & 15.3 & 16.1 & -99.3 & 3543.3 & 2.9\\
5 & -10.3 & 3.6 & -8.1 & 35.3 & -227.8 & 3638.2 & 4.2\\
 \toprule[1mm]
\end{tabular*}
\end{table}

\input{figures}

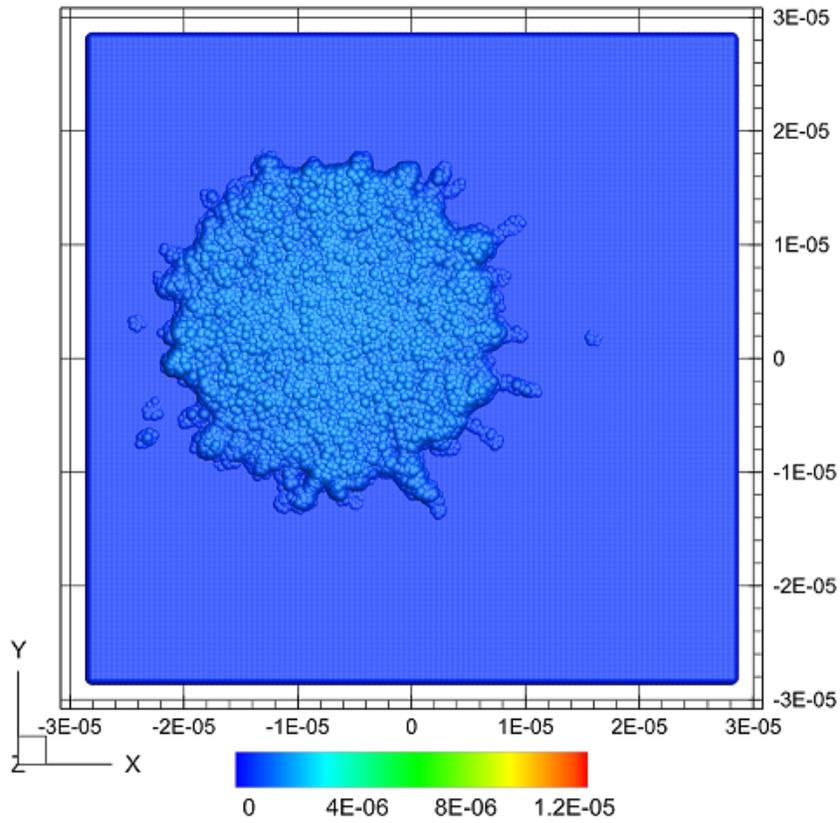
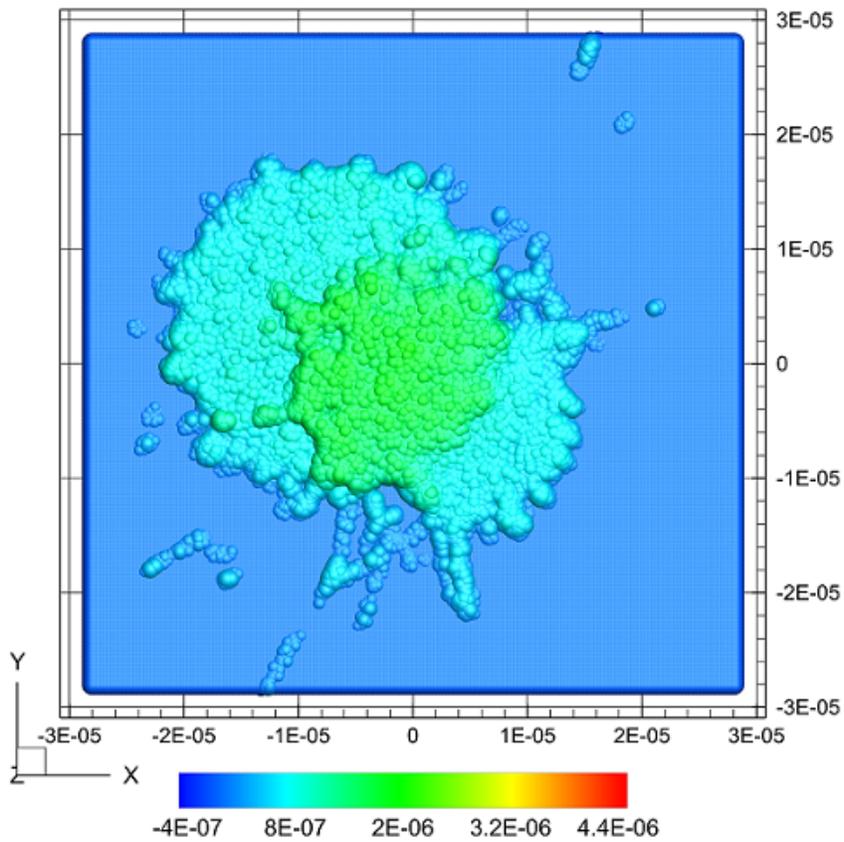
\end{document}